\documentclass{PoS}
\usepackage{textpos}
\usepackage{tikz}
\usepackage{lmodern}
\usepackage{amsmath}
\usepackage{graphics}
\usepackage[skip=-15pt]{caption}
\usepackage{graphicx}
\usepackage{slashed}
\usepackage{mathtools}
\usepackage[thicklines]{cancel}
\usepackage{bm}
\usepackage{xcolor}
\usepackage[document]{ragged2e}
\usepackage[most]{tcolorbox}
\usepackage{subfigure}
\usepackage{adjustbox}
\usepackage{mdframed}
\usepackage{textpos}
\usepackage{framed,color}

\let\OLDthebibliography\thebibliography
\renewcommand\thebibliography[1]{
  \OLDthebibliography{#1}
  \setlength{\parskip}{0pt}
  \setlength{\itemsep}{0pt plus 0.1ex}
}

\newcommand{\U}{{\rm U}}
\newcommand{\SU}{{\rm SU}}
\newcommand{\len}{{\rm len}}

\newcommand{\tr}{\ensuremath{\mathrm{Tr}}}
\makeatletter
\newcommand{\raisemath}[1]{\mathpalette{\raisem@th{#1}}}
\newcommand{\raisem@th}[3]{\raisebox{#1}{$#2#3$}}
\makeatother

\newcommand{\ijkll}{_{\bm{i\,\raisemath{3pt}{j},k\,\raisemath{3pt}{l}}}}

\newcommand{\aibjkll}{_{m+\bm{i\,}\raisemath{3pt}{n+\bm{j}},\bm{k\,\raisemath{3pt}{l}}}}

\newcommand{\ji}[1]{{}_{j_{#1}}^{\;\;i_{#1}}}

\newcommand{\lk}[1]{{}_{l_{#1}}^{\;\;k_{#1}}}

\newcommand{\parfrac}[1]{\frac{\partial}{\partial #1}}
\title{Towards a Dual Representation of Lattice QCD}

\ShortTitle{Towards a Dual Representation of Lattice QCD}

\author{\speaker{Giuseppe Gagliardi}\\
        Fakult\"at f\"ur Physik, Universit\"at Bielefeld, D-33615 Bielefeld, Germany\\
        E-mail: \email{giuseppe@physik.uni-bielefeld.de}}

\author{Wolfgang Unger\\
        Fakult\"at f\"ur Physik, Universit\"at Bielefeld, D-33615 Bielefeld, Germany\\
        E-mail: \email{wunger@physik.uni-bielefeld.de}}

\abstract{Our knowledge about the QCD phase diagram at finite baryon chemical potential $\mu_{B}$ is limited by the well known \textit{sign problem}. The path integral measure, in the standard determinantal approach, becomes complex at finite $\mu_{B}$ so that standard Monte Carlo techniques cannot be directly applied. As the sign problem is representation dependent, by a suitable choice of the fundamental degrees of freedom that parameterize the partition function, it can get mild enough so that reweighting techniques can be used. A successful formulation, capable to tame the sign problem, is known since decades in the limiting case $\beta\to 0$, where performing the gauge integration first, gives rise to a dual formulation in terms of color singlets (MDP formulation). Going beyond the strong coupling limit represents a serious challenge as the gauge integrals involved in the computation are only partially known analytically and become strongly coupled for $\beta>0$. We will present explict formulae for all the integral relevant for $\SU(N)$ gauge theories discretised \`a la Wilson, and will discuss how they can be used to obtain a positive dual formulation, valid for all $\beta$, for pure Yang Mills theory. }

\FullConference{The 36th Annual International Symposium on Lattice Field Theory - LATTICE2018\\
		22-28 July, 2018\\
		Michigan State University, East Lansing, Michigan, USA.}

\begin{document}

\section{Introduction}
Lattice QCD at finite baryon-chemical potential is affected by the \textit{sign problem}. At non-zero $\mu_{B}$ the LQCD action becomes complex giving rise to an exponentially hard problem. Although various techniques have been developed in the past decades in order to circumvent the sign problem, an ultimate solution is still lacking. 
A promising approach that we want to discuss here is the dual variables approach. The key point in this approach is realising that the sign problem is representation dependent. This means that by a suitable change of the degrees of freedom, it is possibile to write down the partition function in terms of states that are closer to the true eigenstate of the Hamiltonian, resulting in a much milder sign problem. Dual formulations have been used in the past years to alleviate, or even solve, the sign problem in various model (see for instance \cite{Vairinhos:2014uxa}-\cite{Marchis:2017oqi}). 
Here we want to discuss the dual approach in Yang-Mills theory and in full QCD from the perspective of strong coupling expansion. 
At $\beta=0$, corresponding to the strong coupling limit, the partition function can be written in terms of dual (integer) degrees of freedom representing mesons and baryons \cite{Rossi_Wolff}. This dual formulation has the advantage that the sign problem induced by a baryon-chemical potential is mild enough so that the phase boundaries can be mapped out using standard reweighting in the sign. Incorporating leading order $\beta$-correction is also possible, by computing the modified  weights induced by a single plaquette excitation \cite{deForcrand:2013ufa}, whereas in \cite{Gagliardi:2017uag} gauge contributions produced by plaquette-surfaces have been taken into account. Going beyond these approximations is very challenging. First of all, the link integrals that appear are not completely known for $\SU(N)$. In addition, a plaquette induced sign problem can appear at $\beta > 0$, limiting the applicability of this method to small $\beta$ values. We will discuss these issues. In particular, we will solve the problem of link integration finding explicit formulae for polynomial integrals over $\SU(N)$. After analysing the sign problem in SC-LQCD with plaquette surface excitations, we will focus on Yang-Mills theory finding a dual, positive, representation by integrating out the gauge links.     

\section{Formulation and Link Integration}
In the following we will always consider the standard LQCD partition function with gauge action discretised \`a la Wilson and 1 flavour of unimproved staggered quarks:
\begin{eqnarray}
\label{Z_standard}
\mathcal{Z} &=& \prod\limits_{x}\int d\chi_{x}d\bar{\chi}_{x}e^{2am_{q}\bar{\chi}_{x}\chi_{x}}\prod_{\ell}\int_{G}dU_{\ell}e^{\sum_{p}\frac{\beta}{N}\operatorname{Re}(\tr U_{p})}\cdot e^{\tr \left[U_{\ell}\mathcal{M}_{\ell}^{\dag} + U^{\dag}_{\ell}\mathcal{M}_{\ell}\right]} \nonumber \\
\left(\mathcal{M}^{\dag}\right)_{i}^{j} &=&  \eta_{\mu}(x)e^{a\mu_{B}\delta_{\mu,0}}\bar{\chi}_{x}^{i}\chi_{x+\mu,j},\qquad  \mathcal{M}_{k}^{l} = -\eta_{\mu}(x)e^{-a\mu_{B}\delta_{\mu,0}}\bar{\chi}_{x+\mu}^{k}\chi_{x,l},
\end{eqnarray}
where $(\ell, x, p)$ label lattice links, sites and plaquettes. After performing a strong coupling expansion in $\beta$, Eq.~(\ref{Z_standard}) can be written as:
\begin{eqnarray}
\prod\limits_{x}\int d\chi_{x}d\bar{\chi}_{x}\,\,e^{2am_{q}\bar{\chi}_{x}\chi_{x}}\sum_{\{n_{p},\bar{n}_{p}\} }\!\prod_{\ell,p}\frac{\left(\beta/2N\right)^{n_{p}+\bar{n}_{p}}}{n_{p}!\bar{n}_{p}!}\!\!\int_{G}dU_{\ell}\tr[U_{p}]^{n_{p}}\tr[U^{\dag}_{p}]^{\bar{n}_{p}}e^{\tr \left[U_{\ell}\mathcal{M}_{\ell}^{\dag} + U^{\dag}_{\ell}\mathcal{M}_{\ell}\right]} \nonumber
\end{eqnarray}
and we introduced the new collective variables $\{n_{p},\bar{n}_{p}\}$ called plaquette (anti-plaquette) occupation numbers. As usual for dual formulations, we wish to integrate out some of the original degrees of freedom. In this case we want to get rid of the $U_{\ell}$ links, by explicitly performing the group integration first. Even though this is quite straightforward in the case $\beta=0$ \cite{Eriksson}, plaquette contributions give rise to serious complications. To show this explicitly, let us consider the $O(\beta)$ corrections to SC-LQCD by Taylor expanding the gauge action to first order:
\begin{eqnarray}
\prod\limits_{\ell}\!\!\int_{G} dU_{\ell}\,\,e^{\frac{\beta}{2N}\tr[ U_{p}+U_{p}^{\dag}]}\cdot e^{\tr \left[U_{\ell}\mathcal{M}_{\ell}^{\dag} + U^{\dag}_{\ell}\mathcal{M}_{\ell}\right]} \approx
\prod\limits_{\ell}\!\!\int_{G} dU_{\ell}(1+\underbrace{\frac{\beta}{2N}\tr[U_{p}+U_{p}^{\dag}]}_{O(\beta) \text{correction}})\cdot e^{\tr \left[U_{\ell}\mathcal{M}_{\ell}^{\dag} + U^{\dag}_{\ell}\mathcal{M}_{\ell}\right]} \nonumber
\end{eqnarray}
The relevant $O(\beta)$ contribution, after performing a hopping parameter expansion of the fermionic action, is given by:
\begin{eqnarray}
\label{order_beta}
\prod\limits_{\ell\in \mathcal{C}(p)}\int_{G}  dU_{\ell}\frac{\beta}{2N}\tr\!\!&[&\!\!\!U_{p}]\,\,e^{\tr \left[U_{\ell}\mathcal{M}_{\ell}^{\dag} + U^{\dag}_{\ell}\mathcal{M}_{\ell}\right]} =\tr\bigg[\prod_{\ell\in \mathcal{C}(p)}J_{\ell}\bigg], \quad\,\mathcal{C}(p)= \{(x,\mu) \in \partial p\} \nonumber \\
&(&\!\!\!J_{\ell}[\mathcal{M},\mathcal{M}^{\dag}])_{m}^{n} =\sum\limits_{\kappa_{\ell},\bar{\kappa}_{\ell}}\frac{1}{\kappa_{\ell}!\bar{\kappa}_{\ell}!}\prod_{\alpha=1}^{\kappa_{\ell}}(\mathcal{M}_{\ell})_{j_{\alpha}}^{i_{\alpha}}\prod_{\beta=1}^{\bar{\kappa}_{\ell}}(\mathcal{M}^{\dag}_{\ell})_{l_{\beta}}^{k_{\beta}}\,\,\,\textcolor{red}{\mathcal{I}^{k_{\ell} +1,\bar{k}_{\ell}}\aibjkll}
\end{eqnarray}
where  $\textcolor{red}{\mathcal{I}^{n+1,n}}$ is the polynomial gauge integral that must be computed at this order. The open color indices $\{i,j,k,l\}$ must be saturated with fermionic sources $\mathcal{M},\mathcal{M}^{\dag}$ while $m,n$ are contracted along the countour $\partial p$ of the plaquette $p$ so that color singlets are recovered afterwards. Away from strong coupling, where $O(\beta^{2}),O(\beta^{3}),..$, contributions are important, all the integrals $\mathcal{I}^{a,b}$ will in general appear. Having explicit formulae for these integrals is the first step towards a dual representation of non-abelian gauge theories in a strong coupling expansion framework. 
Their explicit expression is given by:
\begin{equation}
\mathcal{I}^{a,b}\ijkll = \int_{G}dU\prod_{\alpha=1}^{a} U_{i_{\alpha}}^{j_{\alpha}}\prod_{\beta=1}^{b}(U^{\dag})_{k_{\beta}}^{l_{\beta}}
\end{equation}
where $dU$ is the usual invariant Haar measure and depending on the gauge group $G$ the following constraints apply:\footnote{This gives a constraint on the $\{n_{p},\bar{n}_{p},k_{\ell},\bar{k}_{\ell}\}$ that are allowed. For more details see \cite{Gagliardi:2017uag}.}
\begin{equation}
\mathcal{I}^{a,b}\ne 0 \iff \begin{cases}
      a=b \,\,\qquad\quad\,\,\,\,\U(N) \\
      a=b\,\, {\rm mod} \,\, N \,\,\,\, \SU(N)  \\
\end{cases}
\end{equation}
These integrals were studied extensively in the past, mainly in the case $G=\U(N)$, which was completely solved in \cite{2006CMaPh.264..773C}. Creutz \cite{Creutz} found an explicit formula for the generating functional:
\begin{eqnarray}
\mathcal{Z}^{a,b}[K,J] &=& \int_{\SU(N)}dU\tr[UK]^{a}\tr[U^{\dag}J]^{b} \nonumber \\
\mathcal{I}^{a,b}\ijkll &=&\frac{1}{a!b!}\left(\parfrac{K_{j_{1}i_{1}}}\ldots\parfrac{K_{j_{a} i_{a}}}\parfrac{J_{l_{1}k_{1}}}\ldots\parfrac{J_{l_{b}k_{b}}}\mathcal{Z}^{a,b}[K,J])\right)_{K=J=0} 
\end{eqnarray}
for the case $b=0$, whereas recently Zuber computed the generating functional in the case $a=b+N$ \cite{Zuber:2016xme}. We extended their results in order to cover the most general case $\left(a-b = q\cdot N\right)$, that we present here without proof:
\begin{eqnarray}
Z^{a,b}[K,J] &=& \int_{\SU(N)}dU\tr[UK]^{a}\tr [U^{\dag}J]^{b} \nonumber \\
&\overset{\textcolor{red}{n=\min\{a,b\}}}{=}&(qN+n)!\underbrace{\prod_{i=0}^{N_{c}-1}\frac{i!}{(i+q)!}(\det K)^{q}}_{\text{Baryonic contrib.}}\,\,\underbrace{\sum_{\rho \vdash n }\tilde{W}_{g}^{n,q}(\rho, N)t_{\rho}(JK)}_{\text{Mesonic contrib.}} \nonumber \\
\tilde{W}_{g}^{n,q}(\rho, N) &=&\!\!\!\sum\limits_{\substack{\lambda \vdash n \\ \len(\lambda) \leq N}}\!\!\!\frac{1}{(n!)^{2}}\frac{f_{\lambda}^{2}\chi^{\lambda}(\rho)}{D_{\lambda,N+q}},\qquad\quad t_{\rho}(A) = \prod_{\rho_{i}}\tr(A^{\rho_{i}})
\end{eqnarray}
where $\lambda \vdash n$ means that $\lambda$ is an integer partition\footnote{i.e. $\lambda\vdash n=\left[\lambda_{1},...,\lambda_{k}\right]$ with  $\sum\limits_{i=1}^{k}\lambda_{i}=n$ and $\lambda_{1}\geq\lambda_{2}\geq...\geq\lambda_{k} > 0$. $\len(\lambda)=k$.} of size $n$, $\len(\lambda)$ is its length and $D_{\lambda,N}$, $f_{\lambda}$ are respectively the dimension of the irreps of $\SU(N)$ and $S_{n}$ corresponding to partition $\lambda$. Finally, $\chi^{\lambda}(\rho)$ are the standard $S_{n}$ irreducible characters. The generating functional is splitted in two parts: the first, baryonic contribution, arises from a non-zero $q$ and, being a power of a determinant, gives rise to epsilon tensors after differentiating with respect to source $K$. The second part, the mesonic contribution, is written as a sum over integer partitions that select a particular $\SU(N)$ invariant $\left(\tr_{\rho}(JK)\right)$ weighted by the corresponding factor $\tilde{W}_{g}^{n,q}(\rho,N)$. We called these functions $\tilde{W}_{g}^{n,q}$, ''modified Weingarten functions'' as they correspond to a simple generalization of the standard Weingarten functions obtained in \cite{2006CMaPh.264..773C}. They are all class functions of $S_{n}$ as the partition $\rho$ can be identified as a conjugacy class of permutations $[\pi]$ using the cycle decomposition. This result for the generating functional can be directly used to systematically obtain gauge corrections to any order by using:\footnote{$\mathcal{J}^{a,b}$ is a generalisation, to arbitrary high order, of the integral appearing in Eq.~(\ref{order_beta}).}
\begin{eqnarray}
\label{J_ab}
\mathcal{J}^{a,b}\ijkll[\mathcal{M},\mathcal{M}^{\dag}] &=& \sum_{\{i_\alpha,j_\alpha,k_\beta,l_\beta\}}
\left(\prod_{\alpha=1}^{\kappa_{a}}\mathcal{M}\ji{\alpha}\right)\left(\prod_{\beta=1}^{\kappa_{b}}\mathcal{M}^{\dag}\lk{\beta}\right)\mathcal{I}^{a+\kappa_{a},b+\kappa_{b}}\ijkll \nonumber \\
&=& \frac{k_{a}!k_{b}!}{(a+k_{a})!(b+k_{b})!}\frac{\partial^{(a+b)}\mathcal{Z}^{a+\kappa_{a},b+\kappa_{b}}[J,K]}{\partial K_{j_{1}}^{i_{1}}..\partial K_{j_{a}}^{i_{a}}\partial J_{l_{1}}^{k_{1}}..\partial J_{l_{b}}^{k_{b}}}\bigg|_{\overset{K=\mathcal{M}^{\dag}}{J=\mathcal{M}}}
\end{eqnarray}

\section{Sign problem}
\begin{figure}
\centering
\includegraphics[scale=0.86]{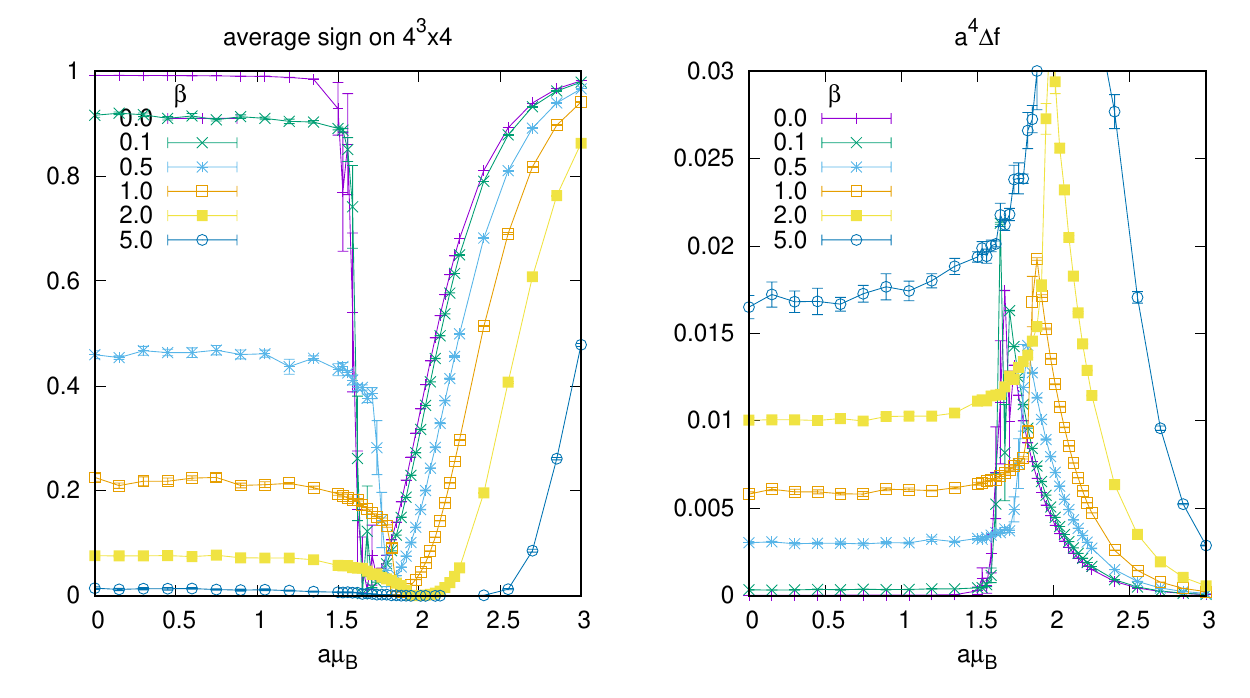} \,\,\,\includegraphics[scale=0.9]{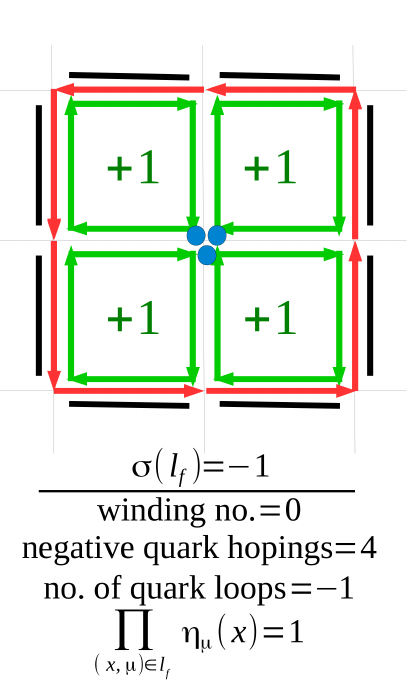}
\vspace{0.7cm}
\caption{\emph{Left}: The average sign $\langle \sigma \rangle= \frac{Z}{Z_{pq}}= e^{-\frac{V}{T}a^{4}\Delta f}$ obtained by simulations on a $4^{3}\times 4$ lattice is shown as a function of the chemical potential $a\mu$ for various $\beta$. \emph{Right}: An example of a configuration with negative sign. An odd number of monomers are trapped in a plaquette surface surrounded by a quark-flux. 
}
\label{Fig:1}
\end{figure}

Although the sign problem is very mild at strong coupling, it could happen that the inclusion of gauge degrees of freedom, in the dual formulation, reintroduce it. This kind of sign problem is absent in the conventional formulation, where a sign problem is only induced by a non-zero $\mu_{B}$. Our result (\ref{J_ab}) for $\mathcal{J}^{a,b}\ijkll$ can be used
to understand how the Monte Carlo weights get modified by plaquette excitations (see \cite{Gagliardi:2017uag}). By making use of the previous result, we performed simulation at finite $\beta$, using an algorithm which is affected by systematic errors only at order $\beta^{N_{c}}$. Results about the phase diagram and details of the simulation can be found in \cite{Gagliardi:2017uag}. Here we want to discuss what happens to the sign problem: In Fig.~\ref{Fig:1} (\emph{Left}), the average sign $\langle\sigma\rangle$ is plotted as a function of $\mu_{B}$ for various $\beta$. The sign problem seems to be immediately reintroduced. Reweighting can be applied only for $\beta < 1$, spoiling the possibility of making contact with the low coupling branch. A sign problem is also present at $\mu_{B}=0$, making it clear that the gauge degrees of freedom alone, as they appear in the dual formulation, produce negative weights. From a diagrammatic point of view, an example of a configuration with negative sign is shown in Fig.~\ref{Fig:1} (\emph{Right}). These findings suggest that to go beyond $\beta=1$, we must study first pure Yang-Mills theory. In particular, we worked on the problem of finding a positive (dual) representation valid for all $\beta$, which will be the topic of the next section. 
\section{Dualization of pure Yang Mills theory}
Let us consider the partition function for pure Yang-Mills theory and expand it in Taylor series around $\beta=0$:
\begin{eqnarray}
\label{Yang_exp}
\mathcal{Z}_{Y.M.} = \sum_{\{n_{p},\bar{n}_{p}\}}\frac{\left(\beta/2N\right)^{\sum_{p}n_{p}+\bar{n}_{p}}}{\prod\limits_{p}n_{p}!\bar{n}_{p}!}\underbrace{\prod_{\ell}\prod_{p}\int_{\SU(N)}dU_{\ell}\left(\tr U_{p}\right)^{n_{p}}\left(\tr U_{p}^{\dag}\right)^{\bar{n}_{p}}}_{\textcolor{red}{\mathbf{W\left(\{n_{p},\bar{n}_{p}\}\right)}}}
\end{eqnarray}
To successfully dualize the partition function, we must find a way to integrate out the gauge fields $U_{\ell}$, expressing the quantity $W\left(\{n_{p},\bar{n}_{p}\}\right)$ in terms of auxiliary degrees of freedom. One way to do this is by decomposing the underlying $\mathcal{I}^{a,b}$ integrals by making use of the Weingarten functions as follows:\footnote{For simplicity we illustrate the procedure in the $\U(N)$ case.}
\begin{eqnarray}
\label{Yang_Wein}
\mathcal{I}^{n,n}\ijkll &=&\!\!\!\!\! \sum\limits_{\sigma,\tau \in S_{n}}\!\!\!\tilde{W}_{g}^{n,0}(\big[\sigma\circ\tau^{-1}\big],N)\delta_{i}^{l_{\sigma}}\delta_{k_{\tau}}^{j} \nonumber \\
\Rightarrow\quad W\left(\{n_{p},\bar{n}_{p}\}\right)&=& \sum_{ \{\sigma_{\ell},\tau_{\ell} \in S_{d_{\ell}}\}}\prod_{\ell}\underbrace{\tilde{W}_{g}^{d_{\ell},0}(\big[\sigma_{\ell}\circ\tau_{\ell}^{-1}\big],N)}_{\gtrless 0}\prod_{x}N^{\boldsymbol\len\big(\big[\hat{\sigma}_{x}\circ\hat{\pi}_{x}\big]\big)}
\end{eqnarray} 
This procedure trades the coloured gauge links $U_{\ell}$, with pair of permutations $(\sigma_{\ell},\tau_{\ell})$. The size of each permutation is determined by the dimer number $d_{\ell}$, defined by:
\begin{eqnarray}
d_{\ell=(x,\mu)} := \min
\begin{cases}
\quad\sum_{\nu > \mu} n_{x,\mu,\nu} + \bar{n}_{x-\nu,\mu,\nu} \\
\quad\sum_{\nu > \mu} \bar{n}_{x,\mu,\nu} + n_{x-\nu,\mu,\nu} \\
\end{cases}\!\Bigg\} 
\end{eqnarray}
The open color indices of the delta functions appearing in Eq.~(\ref{Yang_Wein}) are saturated along the plaquettes to reproduce the traces in Eq.~(\ref{Yang_exp}). This gives rise to powers of $N$. In Eq.~(\ref{Yang_Wein}), the permutation $\hat{\sigma}_{x}$ depends only on $\{n_{p},\bar{n}_{p}\}$ and tells us how the colour flux is re-oriented at each lattice site, while $\hat{\pi}_{x}$ permutes the colour flux on the links attached to $x$ and is defined by:
\begin{equation}
\hat{\pi}_{x} = \bigotimes\limits_{\mu=0}^{d-1}\big(\sigma_{(x,\mu)}\otimes\tau_{(x-\mu,\mu)}\big)
\end{equation}
then $\len\big(\big[\hat{\sigma}_{x}\circ\hat{\pi}_{x}\big]\big)$ is the number of colour cycles at site $x$. This formulation, as it stands, is not suitable for Monte Carlo simulations as almost half of the Weingarten functions appearing in Eq.~(\ref{Yang_Wein}) are negative (see \cite{Gagliardi:2017uag}). Neverthless, it turned out that is possible to rearrange the terms in Eq.~(\ref{Yang_Wein}) in such a way that $W\{n_{p},\bar{n}_{p}\}$ is written as a positive sum:
\begin{eqnarray}
\label{rewriting_W}
W\left(\{n_{p},\bar{n}_{p}\}\right)=
\sum\limits_{\underset{\len(\lambda_{\ell})\leq N}{ \{\lambda_{\ell}\vdash d_{\ell} \} }}\sum_{ \{\sigma_{\ell},\tau_{\ell} \in S_{d_{\ell}}\}}\prod_{\ell}\frac{1}{d_{\ell}!^{2}}\frac{f_{\lambda_{\ell}}^{2}M_{\lambda_{\ell}}^{a_{\ell},b_{\ell}}(\sigma_{\ell})M_{\lambda_{\ell}}^{b_{\ell},a_{\ell}}(\tau^{-1}_{\ell})}{D_{\lambda_{\ell},N}}\prod_{x}N^{\boldsymbol\len\big(\big[\hat{\sigma}_{x}\circ\hat{\pi}_{x}\big]\big)}
\end{eqnarray}
where $M_{\lambda}^{a,b}(\pi)$ is a matrix representation of the irrep $\lambda$ of $S_{n}$ and $a,b =1,..., f_{\lambda}$. We will choose $M$ to be orthogonal matrices\footnote{As $S_{n}$ is a finite group we can  always choose unitary irreps. For the specific case of $S_{n}$ it turns out that the matrix elements are also real.}. After working out the sum over permutations, Eq.~(\ref{rewriting_W}) can be cast in the following form:
\begin{eqnarray}
\label{final_form}
W\left(\{n_{p},\bar{n}_{p}\}\right)&=&\sum\limits_{\underset{\len(\lambda_{\ell})\leq N}{ \{\lambda_{\ell}\vdash d_{\ell} \} }}\overbrace{\left[ \sum_{a_{\ell},b_{\ell}}\prod_{\ell}\frac{1}{D_{\lambda_{\ell},N}}\prod_{x}w(x)\right]}^{W(\{n_{p},\bar{n}_{p}\}, \{\lambda_{\ell}\})\geq 0} \\
P_{\lambda\vdash n}^{a,b} = \frac{f_{\lambda}}{n!}\sum\limits_{\pi \in S_{n}}M_{\lambda}^{a,b}(\pi)\delta_{\pi}, \qquad
w(x) &=& \langle\bigotimes_{\substack{\mu=0 \\ \ell=(x,\pm\hat{\mu})}}^{d-1}P_{\lambda_{\ell}\vdash d_{\ell}}^{a_{\ell},b_{\ell}}, \delta_{\hat{\sigma}_{x}} \rangle_{n_{x}},\qquad n_{x}= \sum\limits_{\mu=0}^{d-1}(d_{x,\mu}+d_{x-\mu,\mu}) \nonumber
\end{eqnarray}
where $\left\{P_{\lambda\vdash n}^{a,b}\right\}$ are a complete set of orthogonal operators in $(\mathbb{C}^{N})^{\otimes n}$ and the inner product $\langle . \rangle_{n}$ is defined as:
\begin{equation}
\langle A,B\rangle_{n} := \tr(A^{\dag}B)\qquad\qquad A,B \in \text{End}\big( (\mathbb{C}^{N})^{\otimes n}\big).
\end{equation}
Therefore, by adding partitions $\lambda_{\ell}$ as an auxiliary degree of freedom, we end up with a partition function that contains only positive terms. However, the possibility of performing Monte Carlo simulations using Eqs.~(\ref{Yang_exp}), (\ref{final_form}), depends on how fast we can compute the weights in Eq.~(\ref{final_form}). Each term involves a sum over $\prod_{\ell}f_{\lambda_{\ell}}^{2}$ local quantities making a brute force computation infeasible in $d>2$. To overcome this issue one possible strategy is to tabularize the weights (as they are $\beta$-independent) or to make use of Tensor Network methods to speed up the computation, which we plan to do in the future.
\section{Conclusion}
We have studied dualization in QCD and in pure Yang-Mills theory from the point of view of the strong coupling expansion. We have solved the problem of computing polynomial integrals over $\SU(N)$ which appear in the procedure of integrating out the gauge links. We showed that plaquette excitations in a na\'ive strong coupling expansion of the gauge action, produce a strong sign problem which limits the use of reweighting to $\beta<1$. As this kind of sign problem is induced by a non-zero $\beta$, we focused on pure Yang-Mills theory, finding a basis where the gluon dynamics does not give rise to a sign problem. This dual basis, where the states are labelled by $\{n_{p},\bar{n}_{p}\}$ and by integer partitions $\lambda_{\ell}$, can reduce the sign problem in full QCD at finite $\beta$.
\section{Acknowledgement}
We acknowledge support by the Deutsche Forschungsgemeinschaft (DFG) through the Emmy Noether Program under Grant No. UN 370/1 and through the Grant No. CRC-TR 211  \textquotedbl Strong-interaction matter under extreme conditions\textquotedbl .


\begin{thebibliography}{99}
\bibitem{Vairinhos:2014uxa}
  H.~Vairinhos and P.~de Forcrand,
  JHEP {\bf 1412} (2014) 038
  [arXiv:1409.8442 [hep-lat]].
\bibitem{Gattringer:2015baa}
  C.~Gattringer, T.~Kloiber and M.~M\"uller-Preussker,
  Phys.\ Rev.\ D {\bf 92} (2015) no.11,  114508
  [arXiv:1508.00681 [hep-lat]].
\bibitem{Marchis:2017oqi}
  C.~Marchis and C.~Gattringer,
  Phys.\ Rev.\ D {\bf 97} (2018) no.3,  034508
  [arXiv:1712.07546 [hep-lat]].
\bibitem{Rossi_Wolff}
P.~Rossi and U.~Wolff, Nucl. Phys. B
248
(1984) 105.
\bibitem{deForcrand:2013ufa}
  P.~de Forcrand, J.~Langelage, O.~Philipsen and W.~Unger,
  PoS LATTICE {\bf 2013} (2014) 142
  [arXiv:1312.0589 [hep-lat]].
\bibitem{Eriksson}
K.~E.~Eriksson, N.~Svartholm  and B.~S.~Skagerstam,
J.~Math.~Phys.~ {\bf 22} (1981) 2276.
\bibitem{Gagliardi:2017uag}
  G.~Gagliardi, J.~Kim and W.~Unger,
  EPJ Web Conf.\  {\bf 175} (2018) 07047
  [arXiv:1710.07564 [hep-lat]].
\bibitem{2006CMaPh.264..773C} 
B.~Collins, P.~{\'S}niady, \ 
2006, Communications in Mathematical Physics, 264, 773. 
\bibitem{Creutz}
M. Creutz, 
J. Math. Phys. {\bf 19} (1978), 2043.
\bibitem{Zuber:2016xme}
  J.~B.~Zuber,
  J.\ Phys.\ A {\bf 50} (2017) no.1,  015203
  [arXiv:1611.00236 [math-ph]].







\end{thebibliography}
\end{document}